\documentclass[twocolumn,preprintnumbers,amssymb,amsmath,9pt,superscriptaddress]{revtex4}[10pt]
\usepackage{graphicx}
\usepackage{dcolumn}
\usepackage{bm}
\begin{document} 
\input epsf.tex
\newcommand{\beq}{\begin{eqnarray}}
\newcommand{\eeq}{\end{eqnarray}}
\newcommand{\nn}{\nonumber}
\def\ltap{\ \raise.3ex\hbox{$<$\kern-.75em\lower1ex\hbox{$\sim$}}\ }
\def\gtap{\ \raise.3ex\hbox{$>$\kern-.75em\lower1ex\hbox{$\sim$}}\ }
\def\CO{{\cal O}}
\def\CL{{\cal L}}
\def\CM{{\cal M}}
\def\tr{{\rm\ Tr}}
\def\CO{{\cal O}}
\def\CL{{\cal L}}
\def\CM{{\cal M}}
\def\mpl{M_{\rm Pl}}
\newcommand{\bel}[1]{\be\label{#1}}
\def\al{\alpha}
\def\bt{\beta}
\def\eps{\epsilon}
\def\eg{{\it e.g.}}
\def\ie{{\it i.e.}}
\def\mn{{\mu\nu}}
\newcommand{\rep}[1]{{\bf #1}}
\def\be{\begin{equation}}
\def\ee{\end{equation}}
\def\bea{\begin{eqnarray}}
\def\eea{\end{eqnarray}}
\newcommand{\eref}[1]{(\ref{#1})}
\newcommand{\Eref}[1]{Eq.~(\ref{#1})}
\newcommand{\gsim}{ \mathop{}_{\textstyle \sim}^{\textstyle >} }
\newcommand{\lsim}{ \mathop{}_{\textstyle \sim}^{\textstyle <} }
\newcommand{\vev}[1]{ \left\langle {#1} \right\rangle }
\newcommand{\bra}[1]{ \langle {#1} | }
\newcommand{\ket}[1]{ | {#1} \rangle }
\newcommand{\fb}{\,{\rm fb}^{-1}}
\newcommand{\ev}{{\rm eV}}
\newcommand{\kev}{{\rm keV}}
\newcommand{\Mev}{{\rm MeV}}
\newcommand{\gev}{{\rm GeV}}
\newcommand{\tev}{{\rm TeV}}
\newcommand{\mev}{{\rm MeV}}
\newcommand{\meV}{{\rm meV}}
\newcommand{\mnu}{\ensuremath{m_\nu}}
\newcommand{\nnu}{\ensuremath{n_\nu}}
\newcommand{\mlr}{\ensuremath{m_{lr}}}
\newcommand{\acc}{\ensuremath{{\cal A}}}
\newcommand{\mav}{MaVaNs}
\newcommand{\disc}[1]{{\bf #1}} 
\newcommand{\mh}{{m_h}}
\newcommand{\hb}{{\cal \bar H}}

\title{Visible Cascade Higgs Decays to Four Photons at Hadron Colliders}
\author{Spencer Chang}
\affiliation{Center for Cosmology and Particle Physics,
  Dept. of Physics, New York University,
New York, NY 10003}
\author{Patrick J. Fox}
\affiliation{Theoretical Physics Group, Lawrence Berkeley National Laboratory,
           Berkeley, CA 94720}
\author{Neal Weiner}
\affiliation{Center for Cosmology and Particle Physics,
  Dept. of Physics, New York University,
New York, NY 10003}
\preprint{}
\date{\today}
\begin{abstract}

The presence of a new singlet scalar particle $a$ can open up new decay channels for the Higgs boson, through cascades of the form $h \to 2a\to X$, possibly making discovery through standard model channels impossible. If $a$ is CP-odd, its decay products are particularly sensitive to physics beyond the standard model. Quantum effects from heavy fields can naturally make gluonic decay, $a \to 2g$, the dominant decay mode, resulting in a $h \to 4 g$ decay which is difficult to observe at hadron colliders, and is allowed by LEP for $\mh > 82 \ \gev$. However, there are usually associated decays with photons, either $h\to 2g\, 2\gamma$ or $h\to 4\gamma$, which are more promising.  The decay $h \to 2 g\, 2 \gamma$ only allows discovery of the $a$ particle and not the Higgs whereas $h \to 4 \gamma$ is a clean channel that can discover both particles.  We determine what branching ratios are required for discovery at LHC and find that with $300 \fb$ of luminosity, a branching ratio of order $10^{-4}$ is sufficient for a large region of Higgs masses.  
Due to a lower expected luminosity of $\sim 8 \fb$, discovery at the Tevatron requires more than $5\times 10^{-3}$ in branching ratio.
\end{abstract}

\maketitle

\section{Introduction}
One of the most important questions in particle physics is the nature of electroweak symmetry breaking. Within the context of the standard model (SM), this is achieved through the vacuum expectation value of the Higgs field. 
The Higgs boson - an excitation about this vev - has well defined couplings, and has important effects on precision electroweak observables through radiative corrections. Current precision electroweak measurements place the best fit value of the Higgs mass to be $\mh = 88 \,\gev$ with an upper bound of $\mh < 146 \,\gev$ at 90\% CL \cite{Erler:2006mt,lepewwg}. In contrast, a SM Higgs boson is already excluded by LEP up to $114.4\,\gev$ \cite{Barate:2003sz}. 

The search for the Higgs must be considered as the cornerstone of any search for weak-scale physics beyond the SM. Because of the strong predictions of the SM, we have many channels in which to search, for instance $h\to bb$ at LEP, $h\to \tau \tau$, $h\to \gamma \gamma$ and $h \to W^+ W^-$ at the LHC. However, in extensions of the SM, it is possible that other decays exist which may dramatically suppress the branching ratios to the expected channels. Such decays could even open up regions of the Higgs mass parameter space which are excluded in the SM, and even make future LHC searches impossible in SM channels.  

Explorations of such non-SM decays have motivated additional LEP analyses, for instance searches for invisible or light jet (flavor independent) Higgs decays.  However, the limits are still quite strong \cite{Achard:2004cf, Achard:2003ty}.  A more recently considered alternative are cascade Higgs decays where the Higgs decays into pairs of a new scalar, $a$, which then subsequently decay \cite{Dobrescu:2000jt}. Neutral singlets appear frequently in theories beyond the SM, including supersymmetry, extra dimensions, and little Higgs theories.  For instance, these decays can occur in the MSSM, where the Higgs decays into the pseudoscalar Higgs, particularly when CP violation is included \cite{Pilaftsis:1999qt,Carena:2000ks,Carena:2002bb}. Cascade decays in the NMSSM are possible (for e.g. \cite{Dobrescu:2000yn, Ellwanger:2005uu}), usually with four b-jet final states, which formerly allowed for more natural theories \cite{Dermisek:2005ar, Dermisek:2005gg}. These are now excluded for $\mh>110\, \gev$ \cite{Sopczak:2006vn}, nearly as strong as the SM limits.  Decays with taus have weaker constraints \cite{Sopczak:2006vn}; four tau decays are still allowed \cite{Sopczak:2006vn} and could be observable at Tevatron \cite{Graham:2006tr}, while   $2j 2\tau$ could be visible at LHC \cite{Ellwanger:2003jt}.  In addition, there can be exotic cascade decays with even more SM final states \cite{Chang:2005ht, Carpenter:2006hs}.

In this paper, the discovery prospects of an interesting new Higgs cascade decay will be analyzed.
The coupling of the Higgs to the new scalar $a$ preserves a $Z_2$ under which $a\to -a$. Thus, any question of the final state of the cascade decay must invoke the question of what violates this $Z_2$. We will argue that a natural possibility is a coupling to heavy fermions, which, if colored, allow the dominant decay to gluons, $a\to 2g$.  In the context of supersymmetric models, we have recently demonstrated how this scenario can be realized in a natural fashion \cite{Chang:2005ht}.  Such a decay is only presently excluded by the OPAL decay-independent study \cite{Abbiendi:2002qp}, which requires $\mh>82 \ \gev$, and the OPAL low mass CP-odd boson search \cite{Abbiendi:2002in}, which places constraints for  $m_a \lsim 12 \,\gev$ when the Higgs mass is below $86 \,\gev$.  A dedicated LEP analysis should give stronger constraints, but most probably still allow a Higgs in the $90-100$ GeV range \cite{Chang:2005ht}.

In this paper we argue that there is generically an associated decay mode of the Higgs into four photons, which can naturally occur with the right order of magnitude for discovery.  The layout of this letter is as follows: in section \ref{sec:h2a} we will review the details of Higgs to two scalar decays, concentrating on $h\to 2a\to 4\gamma$.
In section \ref{sec:collider} we will discuss the search prospects of these decays at hadron colliders.
The prospects are very good at the LHC, such a Higgs decay can be discovered in a large region of expected parameter space.  On the other hand, discovery at the Tevatron requires optimistically larger branching ratios, values that probably require a Higgs mass above the LEP2 kinematic limit, $\sim 120 \,\gev$.
Finally, in section \ref{sec:conc}, we conclude.

\section{Higgs decays to scalars \label{sec:h2a}}
The introduction of a $Z_2$-odd scalar into the SM immediately forces us to consider the coupling ($v=250\,\gev$)
\be
{\cal L} \supset \frac{c}{2} a^2 \, |H|^2 =\frac{c}{4} a^2\, \left(v+h\right)^2
\ee
Such a coupling induces a width $h\to 2a$ \cite{Dobrescu:2000jt} 
\be
\Gamma_{h\to 2a} = \frac{c^2 v^2}{32 \pi m_h}\left(1-4 \frac{m_a^2}{\mh^2}\right)^{1/2}.
\ee
For $\mh \sim 100\, \gev$, a coupling $c = 0.02$ can induce decays to two $a$ at the same rate as two $b$ jets. Present limits on the branching ratio to $b$'s require $c>.06-.1$ for a Higgs in the range $82\ \gev\lsim  \mh \lsim 90\ \gev$ and $c>.04$ in the range $\mh \gsim 90\ \gev$ \cite{Sopczak:2006vn}. Thus, even rather weak couplings can induce dominant $h\to 2a$ decays, while satisfying SM Higgs search limits.  This coupling also gives a contribution to $m_a^2$ of the size $c v^2/2$.  For the values of $c$ required, this contribution gives $m_a <m_h/2$.  Thus, it is quite natural for the $h\to 2a$ decays to be kinematically open.  For the purposes of this paper, we will focus on Higgs masses below 160 GeV, as onshell $W^+W^-$ decays make it difficult for the scalar decay to dominate.  

\subsection*{Scalar decays \label{sec:a2X}}
Now that we have seen that scalar decays can easily dominate over the $h \to bb$ decay, we must ask the question - what couplings allow $a$ to decay? If the $Z_2$ is exact, the signature will simply be the invisible decay of the Higgs boson, which is strongly constrained.   

One interesting and natural possibility for the $Z_2$ is CP, where the $a$ is a CP-odd (i.e. a pseudoscalar), which is what we will assume for the rest of the paper.  Notice that this forbids mixing with the Higgs boson, through trilinear terms like $\mu\, a |h|^2$. 
In the SM, under CP, no renormalizable couplings allow $a$ to decay, which forces us to consider what couplings might exist when beyond the SM physics is included.  
One possibility is that such a field might mix with a new pseudoscalar Higgs $A_0$, for instance as in a two Higgs doublet model. This would allow $a \to 2b$ and $a \to 2\tau$ decays generically, and is well studied and constrained. If the $A_0$ is heavy, or if the mixing is loop suppressed, these decays could be highly suppressed. Indeed, it is quite possible that such a mixing would not exist at all.  

Another possibility is a coupling to heavy vectorlike fermions. This is an appealing possibility as such fermions frequently occur in beyond the SM theories. For instance, in supersymmetry, the Higgs is vectorlike, and has no symmetry reason to be so light.  So it is natural to expect other vectorlike matter at a comparable scale. In little Higgs theories, new top partners are needed to generate the top Yukawa.  In extra dimensional theories, new vectorlike states often appear at the compactification scale.

Whatever its origin, we consider the coupling 
\be
{\cal L} \supset {\bar \psi}(M+ i \gamma_5 \lambda \, a)\psi,
\ee
where $\psi$ is some new fermion, charged under the SM.
Integrating out $\psi$ gives loop induced couplings to gauge bosons. 
These allow the decay $a \to  2x$, where $x=g,\gamma$, with width
\begin{equation}
\Gamma_i = \frac{9 \lambda^2 b_i^2 \alpha_i^2}{1024 \pi^3 M^2}\, m_a^3 N_D,
\end{equation}
where $N_D$ is the multiplicity factor in the final state (i.e. 1 for 
photons and 8 for gluons) and $b_i$ is the contribution of the vectorlike fermion to the beta function for the given gauge group, $U(1)$ electromagnetism or $SU(3)$ color.  For \eg , if $\psi$ is a vectorlike down quark, $b_{SU(3)} = 2/3$ and $b_{U(1)_{em}} = 4/9$.

If $\psi$ fills out an $SU(5)$ multiplet, for instance a $\mathbf{5}$, 
and both the $h \to 2a$ and $a \to 2x$ decays dominate, we have the branching ratios
\be
Br(h\to 4\gamma) \approx 1.4 \times 10^{-5}, \quad Br(h\to 2g 2\gamma) \approx 7.6 \times 10^{-3} 
\ee
A preliminary analysis suggests that the $h\to 2g 2\gamma$ decay at this rate is visible at the LHC but not the Tevatron \cite{Dobrescu:2000jt}.  However, this decay at the LHC might only discover the $a$ boson, due to the difficulty in measuring and finding the soft jets of the two partonic gluons amongst combinatorial background.  Similarly, another discovery mode for the $a$ is its production through gluon fusion and it's subsequent decay into photons, which could be observable at Tevatron/LHC (through an adapted analysis similar to \cite{Landsberg:2000ht,Mrenna:2000qh}). On the other hand, the $h\to 4\gamma$ decay is a clean channel that would discover both particles, the only question is if the rate is large enough to be detected.  In an attempt to answer this question, we will analyze the detection prospects of this channel at hadron colliders \footnote{We are aware of another Higgs scenario that results in four photon final states \cite{Akeroyd:2005pr}.  In their case, the four photons do not form a mass peak as there is an associated $W$ boson in the decay.  Since they only search for an excess in the inclusive $\gamma \gamma \gamma (\gamma) + X$ channel, it would be interesting to see if they can also reconstruct the masses of their two Higgs scalars via reconstructing the $W$.}.  At the LHC, our preliminary analysis will suggest that the $10^{-5}$ branching ratio is almost sufficient. However, it's also worth noting that there may be incomplete GUT multiplets that contribute only to the photon decays.  A prime example comes from the Higgsinos of supersymmetry which can increase the branching ratio into photons \cite{Chang:2005ht}.  For instance, with a vectorlike $\mathbf{5}$ under $SU(5)$, coupling $a$ to Higgsinos with the same mass and coupling as the $\mathbf{5}$ increases the branching ratio to $1.3 \times 10^{-4}$, with even larger values if the Higgsinos are lighter or more strongly coupled to $a$.  Incidentally, the scalar superpartners do not couple linearly to $a$ and hence do not affect the branching ratio. 

However, we expect that LEP data restricts how large this branching ratio can be, $Br(h\to 4\gamma) \lsim 10^{-3}$, for $m_h \lsim 120 \,\gev$. Above this mass range, a larger branching ratio is allowed, but, as we shall see, even a $10^{-4}$ branching ratio allows discovery over most all of the parameter space.  In contrast, discovery at Tevatron will require branching ratios greater than $5\times 10^{-3}$.  Since we expect these branching ratios to be constrained by LEP, this suggests that Tevatron can only probe Higgs masses above about 120 GeV.  At any rate, we will allow all values of the branching ratio, since depending on the vectorlike fields that $a$ couples to, it can be as large as order one and since we also do not know of any specific numerical LEP limits on this branching ratio.  

Another interesting consequence of this weak $Z_2$ violation is the relatively long lifetime of $a$ which could lead to visible 
displaced vertices.  Taking the dominant $2g$ decays as a rough measure of the decay width, one gets 
\be
c \tau_a \sim \frac{1}{\Gamma_{a\to gg}} = 1 \, {\rm cm} \left(\frac{30 \,\gev}{m_a}\right)^3 \left(\frac{M}{450\, \tev}\right)^2 \left(\frac{0.1}{\lambda\,b_3}\right)^{2}
\ee
Since new physics often appear at these same mass scales (e.g. 
gauge mediated supersymmetry breaking), and because $\lambda$ can be naturally quite small, it is motivated to consider the possibility of these displaced vertices.  A recent study suggests that displaced vertices of sufficient length could enhance the Higgs discovery prospects at Tevatron and LHC \cite{Strassler:2006ri}.  However, for the rest of the paper, we will assume that the $a$ scalar decays promptly enough to not have detectable displacements.  

\section{Detection prospects at colliders \label{sec:collider}}
For a light SM Higgs, one of the important search channels is also a rare decay, $Br(h\to 2\gamma) \sim 10^{-3}$ \cite{unknown:1999fq}.  The background of a $4\gamma$ signal is smaller than that for $2\gamma$ so it is not unreasonable that branching ratios as small as $10^{-(4-5)}$ might be detectable.  It also has the benefit of allowing detection of both scalars; 
as long as the $a$ mass is not too small (i.e. $m_a \gtrsim m_h/40$) \cite{Dobrescu:2000jt}, the $a$ particles are only mildly boosted, keeping the photons separated enough to be experimentally reconstructed as a four photon event.  The next question is if the rate is enough for detection.  We will see that the larger integrated luminosity of the LHC will probe almost all of the parameter space of this decay, whereas the Tevatron will probably only be able to discover Higgses above roughly 120 GeV.  For the rest of this analysis, we will focus on the prospects at the LHC, but will mention what changes for the Tevatron. 

\noindent {\bf Cuts:} To analyze the detection reach at the LHC, we implement a parton-level analysis of the signal and background, leaving a more realistic simulation to future work.  We apply the following cuts on our analysis:
\begin{itemize}
\item{Transverse Momentum: $p_T > 20$ GeV for all photons.}
\item{Isolation: $\Delta R >.4$ between all photons}
\item{Rapidity Acceptance: $|\eta|<2.5$}
\item{Consistent Pairing: Require a photon pairing such that $|m_{\rm pair 1}-m_{\rm pair 2}| < 5 \ \gev$.}
\end{itemize}  
The $p_T$ cut satisfies the triggering requirement and also helps to reduce the background.  It will unfortunately lead to small signal acceptances and we will discuss the issues with lowering this cut later.  The isolation cut leaves the option of a more realistic photon isolation condition, which we do not attempt to implement but would be required in a more detailed study.  The rapidity acceptance cut focuses the analysis in the detector region capable of precision EM measurements.  The consistent pairing is an attempt to veto on background events that are inconsistent with a cascade decay.  For the signal we used CTEQ5L, the default settings for PYTHIA, for the parton distribution functions whereas for the background we used CTEQ6L1 which maximizes rates (and thus makes our analysis more conservative).  To take into account photon identification, we use a photon reconstruction efficiency of $80\%$ per photon \cite{unknown:1999fq}, but we would like to point out that detector simulations would be required to determine if this is realistic; however, even in the worst case, we don't expect this to be smaller than $60\%$.  The only detector effects we take into account are Gaussian smearing of the photon energies and angles with numbers given by ATLAS \cite{unknown:1999fq}.  We do not do so for the background, although we expect that with our relatively weak cuts that the numbers will not be largely affected and should be within our background uncertainty.          

\begin{figure}[t]
\includegraphics[width=8.8cm]{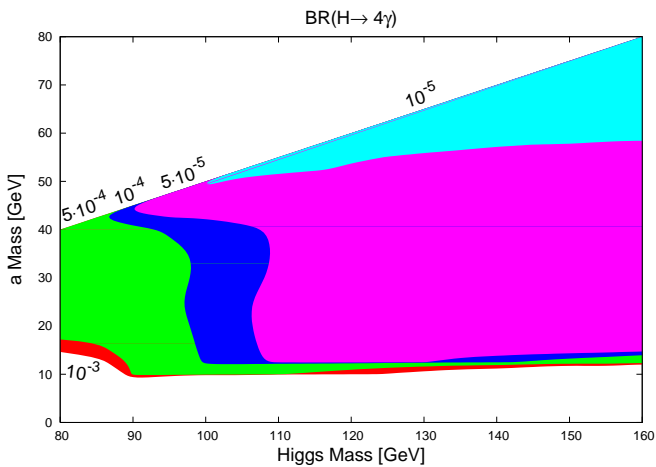}
\caption{Branching ratios sufficient for discovery ($\geq 5$ signal events), given 300 $\fb$ of integrated luminosity at LHC. The bottom region is cut off by the lack of photon isolation, and we consider the region up to $m_h \le 160 \,\gev$, above which we expect $W^+W^-$ decays to dominate. \label{fig:branchingratio}}
\end{figure}

\noindent {\bf Signal: }
For the signal, we implemented into PYTHIA \cite{Sjostrand:2006za} the $h\to 2a$ and $a\to 2\gamma$ decays.  Using this we determined the signal acceptance for the above cuts.  For a fixed $m_h$, the acceptance is much larger when $m_a$ is almost $m_h/2$ and decreases quickly when $m_a$ is below about 10 GeV, this behavior is reflected in the shape of Figure \ref{fig:branchingratio}.  The first effect is due to the $p_T$ cut and the second effect is due to the boosted photon pairs failing the isolation requirement.  
Away from these two extremes, the acceptance increase with Higgs mass which is sub-percent until $100 \,\gev$, about 5\% at 130 GeV, and above 10\% at 160 GeV.  

Within our simulation, we evaluated the expected mass resolutions.  We found that $\Delta m_{h(a)}\sim 0.1\sqrt{m_{h(a)}/\gev}$, $\mathcal{O}(1\gev)$, $\mathcal{O}(0.5\gev)$ respectively, in the region of interest. 
Given the weak consistent pairing requirement, sometimes there are incorrect photon pairings within the signal which give fake $m_a$ solutions.  To distinguish the correct pairing from the incorrect it is usually sufficient to look for the most consistent value of $m_a$; a tighter consistent pairing requirement would also cut down on the potential incorrect pairings.

Finally, to take into account higher order corrections to the production cross section, we use a mass independent K factor of 2, which is characteristic of NNLO calculations in our mass range \cite{Harlander:2002wh,Anastasiou:2002yz}.  In doing so, we assume that the acceptances and mass resolutions would not change much under the NNLO calculation.        

\noindent {\bf Background:}
For the LHC, we used ALPGEN \cite{Mangano:2002ea} to estimate the background both from prompt photon production as well as jets faking photons.  Ultimately, our background estimate will be small enough that order one corrections/refinements will not affect our final results.
To take into account the jet fake rate, we take the numbers given in the ATLAS TDR \cite{unknown:1999fq}, and fit it to a linear function for 20-40 GeV and a constant above 40 GeV, so that for a jet of a given $p_T$, 1 in ${\rm min}(3067,-1333+ 110 p_T/\gev)$ fakes a photon.  We nominally use a factorization scale of $\mu_{p_T} = \sqrt{\sum_{i=1}^n {p_T^2}/n}$ although we also calculate the cross sections for $\mu_F = \mu_{p_T}/2 \; {\rm and} \; 2\mu_{p_T}$, and take the largest result.  In this way, we are being pessimistic on the higher order corrections to the background.  Using ALPGEN, we computed the background of $i \gamma + (4-i) \gamma_j$ (for $i=0-4$), where $\gamma_j$ is a jet faking a photon.  
Despite the cross section for jet production being larger than that for photon production the jet fake rate is small enough that the dominant source of prompt background is 4 photon production.  Since ALPGEN only computes at tree level, there are some processes missed, such as $gg \to 4\gamma$.  However, for the background of Higgs to two photons, it is known that the gluon fusion contribution is about $33\%$ of the tree level processes \cite{Balazs:1998bm}, which for our purposes is small enough to be neglected.  

In addition, there is a comparable background from multiple interactions within a bunch crossing, primarily from two fake photon production occurring twice in the crossing.  Pointing information for the photons could be a discriminant against this particular background, since there could be distinguishable interaction vertices, but we do not attempt to determine its effectiveness. We simulated this pile-up background for the dominant process of $2\gamma_j \oplus 2\gamma_j$, for simplicity we did not simulate pile-up involving $\gamma + \gamma_j$ or $2\gamma$ as these rates are small enough compared to $2\gamma_j$ to be ignored.  

\begin{figure}[t]
\includegraphics[width=8.8cm]{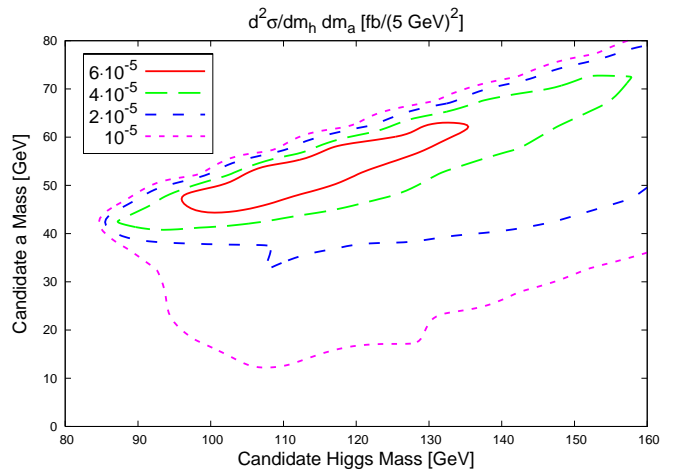}
\caption{LHC background under the cuts given in the text, in femtobarns, binned in 5 GeV windows for both the candidate $m_h$ and $m_a$.  There is no point with value above $10^{-4}$.  \label{fig:background}}
\end{figure}

For an understanding of the size of the background cross section we plot, in Figure \ref{fig:background}, the differential cross section $d^2\sigma/ dm_h dm_a$ for the sum of both the single and multiple interaction backgrounds, as binned in 5 GeV windows for both $m_h$ and $m_a$.  From the simulated mass resolutions of the signal, these mass windows are generously large, but essentially all signal events would be accepted and the background estimate should be robust to the ignored detector effects.  To take into account the possibility of background events with multiple allowed $m_a$, all consistent solutions are included in the plots.  To summarize, we believe our background estimates should be accurate up to factors of order one, and since it is already so small, this uncertainty will not affect our results until we discuss weakening cuts or running at an upgraded higher luminosity mode in the next subsection.      

\noindent {\bf Detection Prospects:} 
Putting all of this together, we can determine what branching ratios are required for discovery given $300 \fb$ at the LHC.  Since the number of background events in a bin (B) is particularly low (with B $\lesssim .03$), $5\sigma$ Poisson statistics would usually require only a couple of events, but to be conservative, we require at least 5 signal events for discovery.  The branching ratios required for this appear in the plot in Figure \ref{fig:branchingratio}.  
From the figure, there appears to be a reasonably large region of parameter space where a branching ratio under $10^{-4}$ is capable of detecting both the Higgs and $a$ scalar.    

In comparison, the Tevatron is sensitive to a smaller region of parameter space.  The background remains small enough to only require 5 signal events to claim discovery.  However, Tevatron's reach is weakened by a smaller integrated luminosity (up to about 8 $\fb$ expected at the end of Run II) and by its lower Higgs production cross section.  In these two factors, Tevatron is down by an order of magnitude each.  On the other hand, since the jet fake rate at Tevatron has been measured down to 10 GeV \footnote{See X. Zhang talk, ``Electron, Muon, and Photon ID at the Tevatron," at the Hadron Collider Physics Symposium 2006.}, we can lower the overall $p_T$ cut to this, which gives more reasonable acceptances (around 10-50\%).  Ultimately, Tevatron requires branching ratios larger than about $5\times 10^{-3}$ to discover the Higgs in this mode.  Since we expect LEP to have strong constraints on such Higgs decays, this suggests that the only range where Tevatron can certainly probe such a Higgs is above about 120 GeV, where LEP limits would not apply.  However, this is strongly dependent on the actual LEP constraint.  If it's weaker then our expected $10^{-3}$ constraint, Tevatron could still probe some Higgses below the LEP2 kinematic limit.         

Even though both colliders are capable of discovering this Higgs for a large region of parameter space, they are not sensitive to the minimal branching ratio of $10^{-5}$ given by the $SU(5)$ example.  For the Tevatron, this is too far out of reach, but at the LHC an order one signal enhancement could be enough.  
With this in mind, we now discuss experimental methods to enhance the signal and suppress the background, in an attempt to increase the reach to smaller values of the branching ratio.

In order to increase signal rates, one either has to increase the acceptance or the integrated luminosity.  The acceptance can be improved, especially for light Higgses, by reducing the overall $p_T$ cut (there still need to be 2 $p_T \geq 20$ GeV photons to trigger on).  In our signal simulation, the acceptance does in fact improve by a factor of 4-10, if the overall $p_T$ cut is lowered to 15 GeV.  Unfortunately, in this case, our background estimation is much less certain.  The fit to the ATLAS jet fake rate is based on their simulations which only go down to 20 GeV.  However, if we extrapolate down to 15 GeV (where 1 in 300 jets fake a photon), our background rates are still under control where B $\lesssim .3$, albeit with large regions where B $\sim .1 -.4$.  In this case, 5 events would still be considered a 5$\sigma$ discovery via Poisson statistics.  Aside from this assumption about the fake rate, we remind you that the other uncertainties in the background (due to unsimulated background, scale dependence), could make the background larger than this estimate and thus it's probably safer to require more than 5 events.  Part of this uncertainty will be reduced, if the jet fake rate can be measured to low enough $E_T$.  For instance, for our analysis at the Tevatron, their measured jet fake rate goes down to 10 GeV, which we used to verify that the background remains under control to such low cuts.  

For the special case of $m_a$ below 10 GeV, acceptance gains would result from replacing our isolation cut with an isolation condition that would allow photons to be closer, this would of course also increase the background rate.  More drastically, acceptance gains would result from changing the search, by looking instead for excesses in inclusive searches for 3 or more photons.  However, there would be no direct way to measure the Higgs mass, making it difficult to claim discovery of the Higgs.  Finally, it is important to emphasize that these acceptances are very sensitive to the photon triggers.  In fact, the signal would be killed if the ATLAS or CMS triggers were pushed much larger than 20 GeV for two photons.  If this is required to meet the trigger rate budget, a multiple photon trigger of 15-20 GeV would still efficiently trigger on this decay.        

In terms of luminosity, LHC or Tevatron experiments could combine their results, giving an additional factor of 2 in the expected number of events.  Looking further ahead, the signal could also benefit from an order of magnitude increase in luminosity, as in an SLHC upgrade \cite{Gianotti:2002xx}; here the background is even less under control, as the multiple interaction background would naively increase by a factor of 100 and thus we would need many more than 5 events.  Therefore, this mode would require stronger background rejection, which will be discussed next.   

There are many ways to lower the background rate, which could be important since our attempts at boosting the signal tend to make the background nonnegligible.  The first thing to gain on would be to tighten up the binning in $m_h$ and $m_a$, since our background plots assumed 5 GeV windows, and given the numbers in the signal section, could be chosen to be as low as 4 and 2.5 GeV respectively (for $\pm 2\sigma$ windows).  Also, the consistent pairing criterion is far too weak, as the signal usually has a mass difference of less than 1 GeV; a tighter cut could make this more efficient at suppressing the background.  In terms of the multiple interaction background, it may also be possible to reject based on discernible multiple interaction vertices.  This is the main issue for going to higher luminosity mode, and to keep the background reasonable would require about an order of magnitude suppression.              

To summarize, we have found that standard cuts render this search essentially background free.  Under the requirement of 5 signal events for discovery, LHC has a wide reach for branching ratios of order $10^{-4}$ and Tevatron has a reach for Higgses heavier than 120 GeV for branching ratios greater than $5\times 10^{-3}$.  For Tevatron, this lower bound on the search of 120 GeV is sensitive to the details of LEP constraints on such Higgs decays and could probe lower Higgs masses if constraints are weaker than we expect.  With some improvements in acceptance or luminosity, the LHC potentially could probe the expected minimal branching ratio $10^{-5}$ for this type of Higgs decay.  

\section{Conclusions \label{sec:conc}}
We have considered the consequences of the introduction of a new pseudoscalar into the Higgs sector of the standard model. Decays of the Higgs into pairs of this pseudoscalar can easily dominate over decays into SM particles until the Higgs is heavy enough to decay into onshell $W$'s. 

However, absent new fields, such a pseudoscalar is stable as no renormalizable coupling in the standard model allows this to decay, making these decays sensitive to physics beyond the standard model. In particular, 
should such a field couple to new, heavy states, decays $a\to 2g$ and $a \to 2\gamma$ are allowed, with widths proportional to $\alpha_s^2$ and $\alpha_{EM}^2$, respectively. At these rates, the decays $h\to 2a\to 2 g 2\gamma$ and $h\to 4\gamma$ are interesting signals. Depending on the masses of the new vectorlike matter, $a$ can propagate macroscopic distances before decaying.

We have considered the required rates for detection of these decays at future hadron colliders.  The $h\to 2g 2\gamma$ decay has been determined to be visible at LHC \cite{Dobrescu:2000jt}, but this and direct production of the $a$ particle might only discover the lighter scalar.  Instead we considered the $h\to 4\gamma$ decay and find that it can be seen at the LHC if the branching ratio is $O(10^{-4})$, which is close to the expected minimal $10^{-5}$ value.   This is a very clean channel that allows discovery of both the Higgs and the $a$ scalar, which could be the only way of discovering the Higgs in these scenarios.  On the other hand, the Tevatron reach is weaker, since it requires branching ratios of $5\times 10^{-3}$, which are subject to LEP constraints.  Therefore, it can probably only probe Higgs masses above about 120 GeV, but this result is dependent on the specific constraints LEP has on such decays.      
Finally, pointing towards future studies of this decay mode at LHC, we have suggested methods to improve the detection prospects, in particular studying the jet fake rate below $20\ \gev$ in order to lower the overall $p_T$ cut, and also pointed out that increasing the threshold of the two photon triggers at ATLAS or CMS could potentially miss this signal, which motivates a multiple photon trigger if the two photon trigger must be tightened.     

\section*{Acknowledgements} We thank Martin Boonekamp, Bob McElrath, Mark Oreglia, Michele Papucci, Maxim Perelstein, Andre Sopczak, Jesse Thaler, Jay Wacker and Scott Willenbrock for useful conversations.  We thank Rikard Enberg for assistance with PYTHIA.  We especially thank Konstantin Matchev and Tilman Plehn for a careful reading of the manuscript.  The work of S. Chang and N. Weiner was supported by NSF CAREER grant PHY-0449818. P.J. Fox was supported in part by the Director, Office of Science, Office of High Energy and Nuclear Physics, Division of High Energy Physics, of the US Department of Energy under contract DE-AC02-05CH11231.

\bibliography{ldecay}
\bibliographystyle{apsrev}

\end{document}